\documentclass[12pt]{iopart}

\usepackage{iopams}
\usepackage{graphicx}
\usepackage{cite}
\usepackage{color}
\usepackage{multirow}
\usepackage{epstopdf}

\begin{document}

\title[Customized IDFs in CCPs by voltage waveform tailoring]{Customized ion flux-energy distribution functions in capacitively coupled plasmas by voltage waveform tailoring}

\author{E. Sch\"{u}ngel$^1$, Z. Donk\'o$^2$, P. Hartmann$^2$, A. Derzsi$^2$, I. Korolov$^2$,  J. Schulze$^1$}

\address{$^1$Department of Physics, West Virginia University, Morgantown, WV 26506, USA\\ $^2$Institute for Solid State Physics and Optics, Wigner Research Centre for Physics, Hungarian Academy of Sciences, 1121 Budapest, Konkoly-Thege Mikl\'os str. 29-33, Hungary.\\ }
\ead{EdSchuengel@mail.wvu.edu}


\begin{abstract}
We propose a method to generate a single peak at a distinct energy in the ion flux-energy distribution function (IDF) at the electrode surfaces in capacitively coupled plasmas. The technique is based on the tailoring of the driving voltage waveform, i.e. adjusting the phases and amplitudes of the applied harmonics, to optimize the accumulation of ions created by charge exchange collisions and their subsequent acceleration by the sheath electric field. The position of the peak (i.e. the ion energy) and the flux of the ions within the peak of the IDF can be controlled in a wide domain by tuning the parameters of the applied RF voltage waveform, allowing optimization of various applications where surface reactions are induced at particular ion energies. 
\end{abstract}


\maketitle

\section{Introduction}

Capacitively coupled plasmas (CCPs) have a wide range of applications, most of which require a solid control of the ion properties, i.e. the ion flux and the mean ion energy, at the electrode (target) surfaces \cite{Makabe_Book,Chabert_Book}. During the past decades different schemes, which allow a separate control of these quantities, have been proposed. Radio frequency discharges driven simultaneously by two significantly different frequencies have approached this issue by the functional separation of the two components of the driving voltage waveform: the high frequency component plays the dominant role for the generation of the plasma and, thereby, sets the ion flux, while the low frequency component is responsible for accelerating the ions towards the electrodes, i.e. it defines the mean energy of the ions at the surfaces \cite{dual1,dual2,dual2a,dual3,dual4,dual5,dual6,fundamental}. A more recent approach, based on the Electrical Asymmetry Effect (EAE) \cite{EAE2,EAE3,EAE5,EAEbienholz,EAE_IRLAS_DEPO,ELIAS3,EAEmultif1,EAEmultif2,EAEjohnson,EAEjohnson2,VWT_Lafleur,Diomede1,Hrunski,fundamental} makes use of the generation of a controllable DC self-bias that develops even in geometrically symmetric reactors when the discharge is excited by the sum of two (or more) consecutive harmonics of a base frequency. The control over the DC self-bias, $\eta$, and over the mean ion energy by changing the phase angles of the applied harmonics has been confirmed both experimentally \cite{EAE5,EAEbienholz,ELIAS3} and in simulations \cite{EAE2,EAE3,ELIAS3,EAEmultif1,EAEmultif2,VWT_Lafleur,Diomede1}, and has been proven benefitial in thin film deposition applications \cite{EAE_IRLAS_DEPO,EAEjohnson,EAEjohnson2,Hrunski}. While these basic approaches and their variants \cite{Shannon} allow controlling these two integral characteristics of the ion flux-energy distribution function (IDF), a control over the {\it shape} of the IDF would be highly desirable, e.g. to drive energy-selective surface processes at a distinct ion energy \cite{Makabe_Book}. For such applications, customizing the IDF to form a narrow peak at controllable energies would be ideal.

Under collisionless conditions, the IDF in a single-frequency CCP typically exhibits a double peak feature at high energies \cite{IEDF1,IEDF2}, whereas collisions lead to a more complex spectrum at lower energies \cite{IEDF3,IEDF4,IEDF5,IEDF6}. Typically, the peaks are formed by 10 \% - 20 \% of all ions flowing to the electrode \cite{fundamental}. So far, a control over the IDF shape has only been realized in high density remote plasma sources with unmatched arbitrary substrate bias \cite{Wendt1,Wendt2,Baloniak,Diomede2} at low repetition rates ($\sim$ 100 kHz) for ion extraction. While this concept is not applicable for large area CCP sources, the several degrees of freedom in the case of EAE-type excitation waveforms may provide a possibility to go beyond the control of the flux and mean energy, i.e. towards the optimization of the complete IDF in CCPs via a specific tuning of the amplitudes and phases of the individual frequency components of the driving voltage. 

The work presented here proposes such a tailoring of the IDF, via utilizing the effect of charge exchange (cx) collisions on the IDF \cite{IEDF3,IEDF4,IEDF5} in CCPs driven by customized voltage waveforms. It will be demonstrated that (i) a narrow peak in the IDF can be obtained under such conditions via adjusting the ion dynamics for our needs, and (ii) the position of this peak (i.e. the energy) and the corresponding ion flux can be controlled via the driving voltage waveform. 
It is of key importance that cx collisions "convert" fast ions to slow ions, which can accumulate in the sheath regions until a high electric field drives them to the electrodes.

We note that at low pressures a narrow ion energy distribution can also be generated in single-frequency CCPs if (nearly) collisionless conditions for the sheaths can be reached \cite{IEDF1,IEDF2}. The energy of the ions in this case is, however, defined by the mean sheath voltage, so it occurs at the high energy end of the IDF. 
These conditions are far from ideal for deposition processes, which require a high voltage (high ion flux) and, at the same time, a peak of the IDF at low/intermediate energies to enhance particular surface reactions. This cannot be realized in classical CCPs, but it is facilitated based on the technique proposed here.

\section{PIC simulation}

We investigate the possibility of the IDF tailoring via self-consistent kinetic simulations based on the Particle-in-Cell technique complemented with Monte Carlo treatment of collision processes (PIC/MCC). We use our 1d3v electrostatic, bounded plasma simulation code\cite{Donko_2011_PSST, fundamental} to study a geometrically symmetric discharge in helium at 6 Pa and 350 K. The plane, parallel, and infinite electrodes are separated by a distance of 5.0 cm. At the electrodes, electrons are reflected with a probability of 0.2 \cite{Kollath_erefl} and secondary electrons due to ion impact are generated with a probability of $\gamma=0.1$. The cross sections for electron-neutral and ion-neutral collision processes are taken from Refs. \cite{He_cs_1,He_cs_2,He_cs_3}. 

Based on the ideas outlined above one can define ideal waveforms, for which the effects will be most pronounced. Keeping in mind possible experimental realizations, we approximate the ideal waveforms with signals consisting of five consecutive harmonics, 
\begin{equation}
\label{AppVol}
\phi_{\rm appl}(t)=A_0 \sum_{k=1}^5 \phi_k \cos(2\pi k f_1 t + \theta_k),
\end{equation} 
that are applied to one of the electrodes via a blocking capacitor, while the other electrode is grounded. Here, $f_1$ = 13.56 MHz. $A_0$ is a voltage factor for all harmonics, while $\phi_k$ and $\theta_k$ are the amplitudes and phases of the individual frequency components. Such a voltage waveform is feasible in applications using an advanced multi-frequency RF supply system with multiple matching branches and electrical filters \cite{Barthel}. Due to the asymmetry of the waveform a DC self-bias develops to equalize the electron and ion fluxes at each electrode, as described by the Electrical Asymmetry Effect \cite{EAE2,EAE3,EAE5,EAEbienholz,EAE_IRLAS_DEPO,ELIAS3,EAEmultif1,EAEjohnson,EAEjohnson2,VWT_Lafleur,Diomede1,Hrunski,fundamental}.
The application of a multi-frequency voltage waveform generally enhances the electron heating \cite{EAEmultif2,VWT_Lafleur}. This leads to a high total ion flux, which is benefitial for most processing applications. 
In the following we present results for two scenarios. First, we investigate the effect of the driving voltage amplitude and, subsequently, we address the effects of modifying the shape of the driving voltage waveform.

\section{Results}

\begin{figure}[t]
\begin{center}
\includegraphics[width=0.75\textwidth]{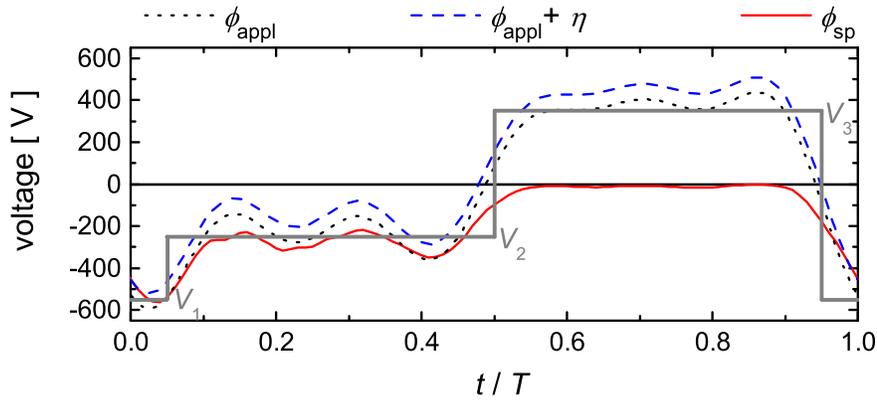}
\caption{Driving voltage waveform ($\phi_{\rm appl}$), discharge voltage including DC self-bias ($\phi_{\rm appl} + \eta$), and voltage drop over the powered electrode sheath ($\phi_{\rm sp}$). The grey lines indicate the ideal applied voltage waveform, consisting of three plateaus with voltage values $V_1$, $V_2$, and $V_3$.}
\label{fig:1}
\end{center}
\end{figure} 

\begin{figure}[t]
\begin{center}
\includegraphics[width=0.75\textwidth]{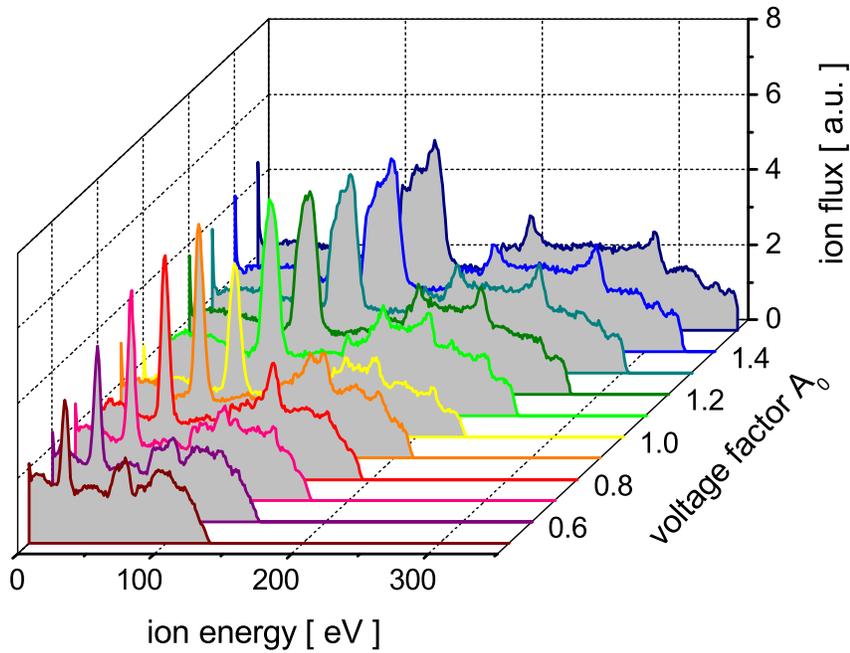}
\caption{Flux-energy distribution functions of He$^+$ ions at the powered electrode, for an excitation voltage waveform according to figure \ref{fig:1}, scaled with different amplitude factors $A_0$.}
\label{fig:2}
\end{center}
\end{figure}

Figure \ref{fig:1} presents the driving voltage waveform, the voltage drop across the discharge gap, and the sheath voltage at the powered electrode as a function of time within one fundamental RF period. The applied voltage waveform is generated by setting $A_0=1.0$ and the individual amplitudes $\phi_1=394$ V,  $\phi_2=143$ V, $\phi_3=153$ V,  $\phi_4=85$ V, $\phi_5=96$ V and phases $\theta_1=107 ^\circ$, $\theta_2=186 ^\circ$, $\theta_3=134 ^\circ$, $\theta_4=195 ^\circ$, $\theta_5=149 ^\circ$ in equation (\ref{AppVol}). These values are obtained from a Fourier transform of the ideal applied voltage waveform. Thus, they have been chosen, because the resulting voltage waveform is minimum for a short time only, then oscillates around an intermediate value and stays close to the maximum value for almost half of the RF period. The idea is that the powered electrode sheath remains collapsed for a long time, so that ions created by cx collisions (which we call "cx-ions" below) within the sheath region can accumulate in front of the electrode. After that, these cx-ions are accelerated by the sheath electric field and arrive at the electrode within one RF period, $T$. This field is strong around $t=0$ and then gradually decreases, so that cx-ions are initially accelerated and are, subsequently, located close to the electrode, where the intermediate field is sufficiently large to ensure that they reach the electrode within one RF period.

\begin{figure}[t]
\begin{center}
\includegraphics[width=0.75\textwidth]{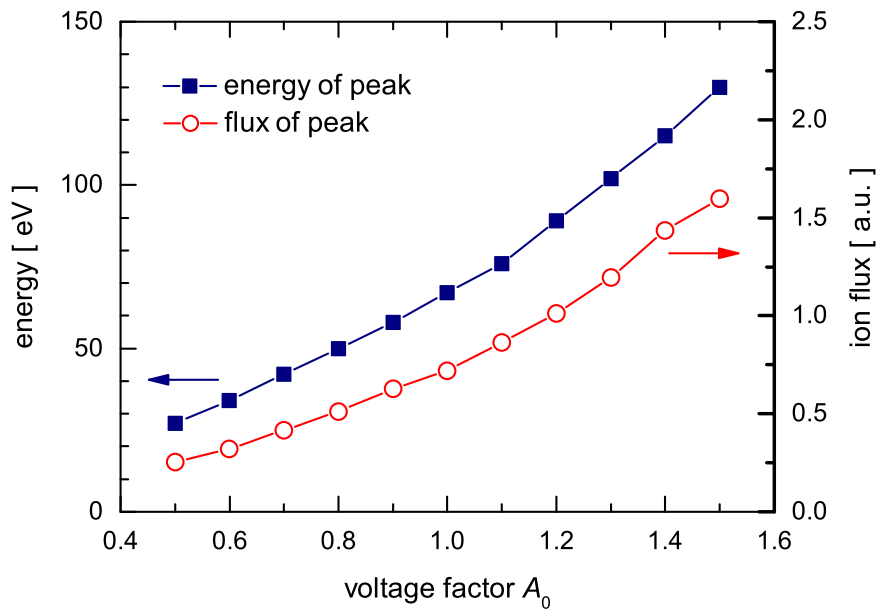}
\caption{Position of the dominant peak in the energy spectrum and the ion flux, obtained as the integral of this peak, as a function of $A_0$.}
\label{fig:3}
\end{center}
\end{figure}
 
As shown in figure \ref{fig:2}, utilizing this concept results in a sharp peak at a distinct ion energy within the low/medium energy range of the IDF. Here, we perform a voltage amplitude variation by changing all the harmonic amplitudes by the same factor, $A_0$, between 0.5 and 1.5, while leaving the phases unchanged. We observe that (i) for all conditions a sharp peak is present in the spectrum, (ii) with changing the voltage factor the position of this peak can be controlled, and (iii) this dominant peak of the IDF broadens with increasing $A_0$. The analysis of the properties of the dominant peak is presented in figure \ref{fig:3}, which shows the energy that corresponds to the peak and the ion flux (integral of the peak). The position of the peak can be tuned within a 27 eV -- 130 eV range by varying the excitation voltage amplitude by a factor of 3.
The flux of the ions that belong to this peak changes by a factor of 6. The relative contribution of the flux of the ions within the peak to the total ion flux also increases, from 15.1 \% at $A_0=$0.5 to 21.1 \% at $A_0=$ 1.5. This is a significant fraction of the total ion flux and comparable or above such fractions achievable by classical techniques in CCPs \cite{fundamental}. However, in contrast to classical techniques our method allows controlling the energy of this peak in the intermediate region of the energy spectrum and generating one single dominant peak. Furthermore, the DC self-bias increases from 27.5 V at $A_0=$ 0.5 to 91.5 V at $A_0=$ 1.5.
The relatively high flux observed in the smallest energy bin (below 1 eV) is caused by cx ions that are created very close to the electrode. A large fraction of these ions gains only very little energy, as the sheath electric field is small for almost the entire second half of the RF period.

\begin{figure}[t]
\begin{center}
\includegraphics[width=0.75\textwidth]{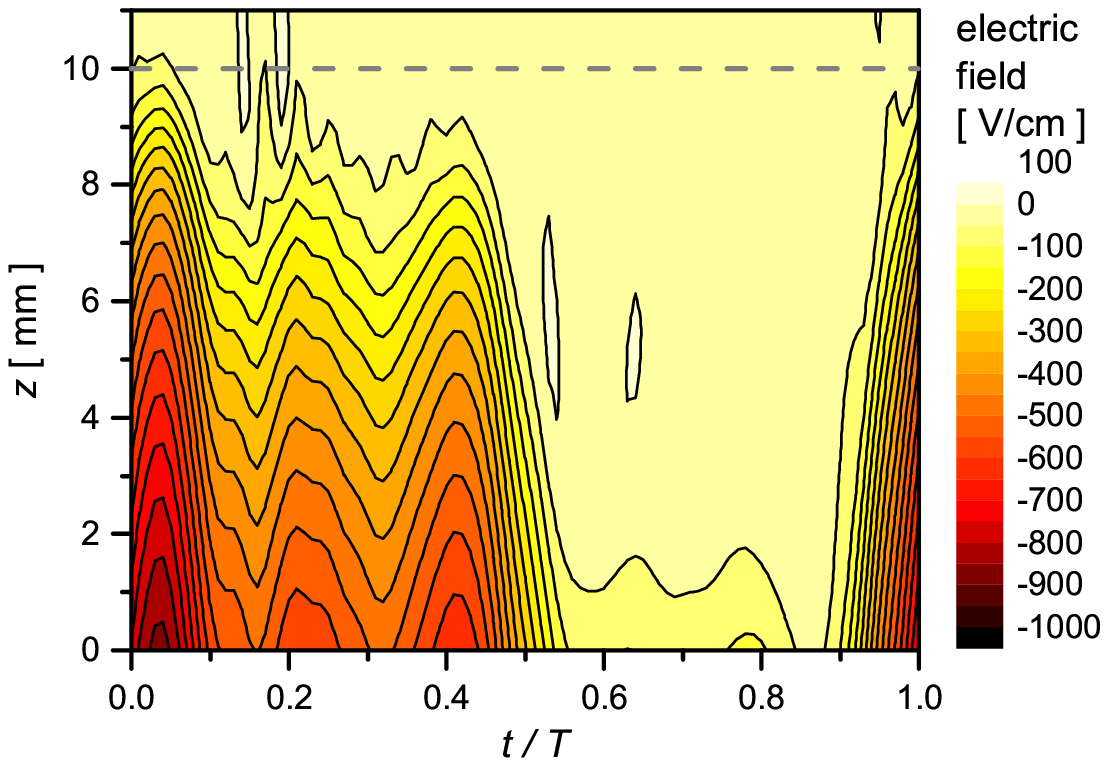}
\caption{Spatio-temporal distribution of the electric field in the powered electrode sheath region. The dashed line at $z\approx10$ mm indicates the maximum sheath extension ($s_{max,p}$). The driving voltage is shown in figure \ref{fig:1} ($A_0=$ 1.0).}
\label{fig:4}
\end{center}
\vspace{13mm}
\begin{center}
\includegraphics[width=0.75\textwidth]{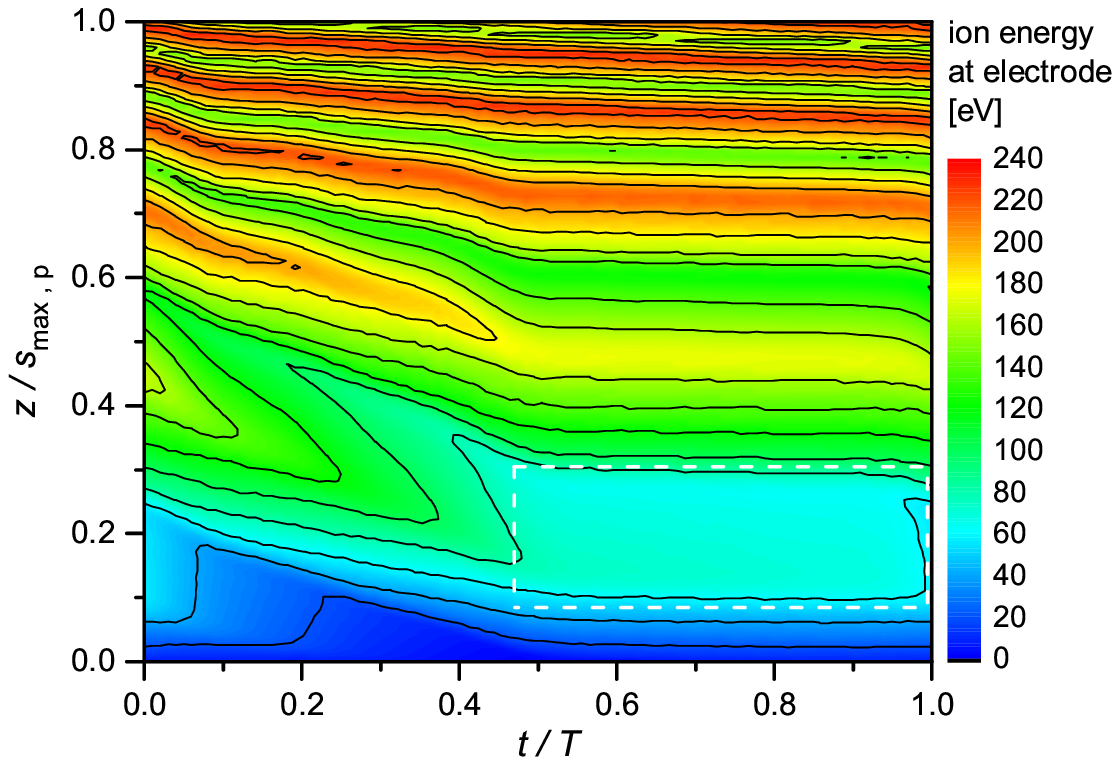}
\caption{Arrival energy of ions as a function of their initial position in space and time. The driving voltage is shown in figure \ref{fig:1} ($A_0=$ 1.0). The white dashed box highlights the region of cx-ion accumulation leading to the peak in the IDF.}
\label{fig:5}
\end{center}
\end{figure}

The understanding of the formation of the main peak in the IDF is aided by analyzing the spatio-temporal distribution of the electric field in the sheath region (see figure \ref{fig:4}) and invoking a method based on the EST (Ensemble-in-Space-Time) \cite{MS} approach, where the trajectories of individual ions are traced. We seed cx-ions, which are initially at rest, into the sheath electric field, which is predefined as a result of the PIC/MCC simulations. We assign the arrival energy of the ions at the electrode (neglecting collisions) to each position in space (between $z=0$ and $z=s_{max,p}$) and time within the sheath region, where the ions started \cite{IEDF5}. This construction is shown in figure \ref{fig:5} for the $A_0=1.0$ case. 
A large domain in this plot, where the ion energy is nearly the same ($\approx65$ eV, marked by the white rectangle), gives rise to the nearly monoenergetic peak in the IDF. This domain establishes because the electric field is strong for a short time, after which it oscillates around an intermediate value and is almost zero between $t=0.5$ $T$ and $t=0.9$ $T$. Thus, cx-ions can accumulate in this time interval close to the electrode. These ions are suddenly accelerated by the following peak in the sheath electric field and arrive at the electrode after gaining energy between $t=0.9$ $T$ and $t=1.5$ $T$. Therefore, all cx-ions created within the window $0.5$ $T \lesssim t \lesssim 0.9$ $T$ and $0.1$ $s_{max,p} \lesssim z \lesssim 0.3$ $s_{max,p}$ arrive with about the same energy, resulting in the distinct peak observed in the IDF. Ions, which start from a position further away from the electrode, need more than one RF period to move through the sheath and, therefore, contribute to the IDF at higher energies; they do not cause another distinct peak due to (i) the lower electric field at the starting position, leading to a much longer transition time, so (ii) the probability of collisions is high, redistributing these ions in the IDF. An increase in the voltage factor simply enhances the sheath electric field and, thereby, enhances both the spatial region, in which ions must be initially located in order to contribute to the peak in the IDF, and the energy of this peak due to the stronger force exerted on the ions. 
In this way the shape and the energy of the peak can be controlled.


\begin{figure}[t]
\begin{center}
\includegraphics[width=0.75\textwidth]{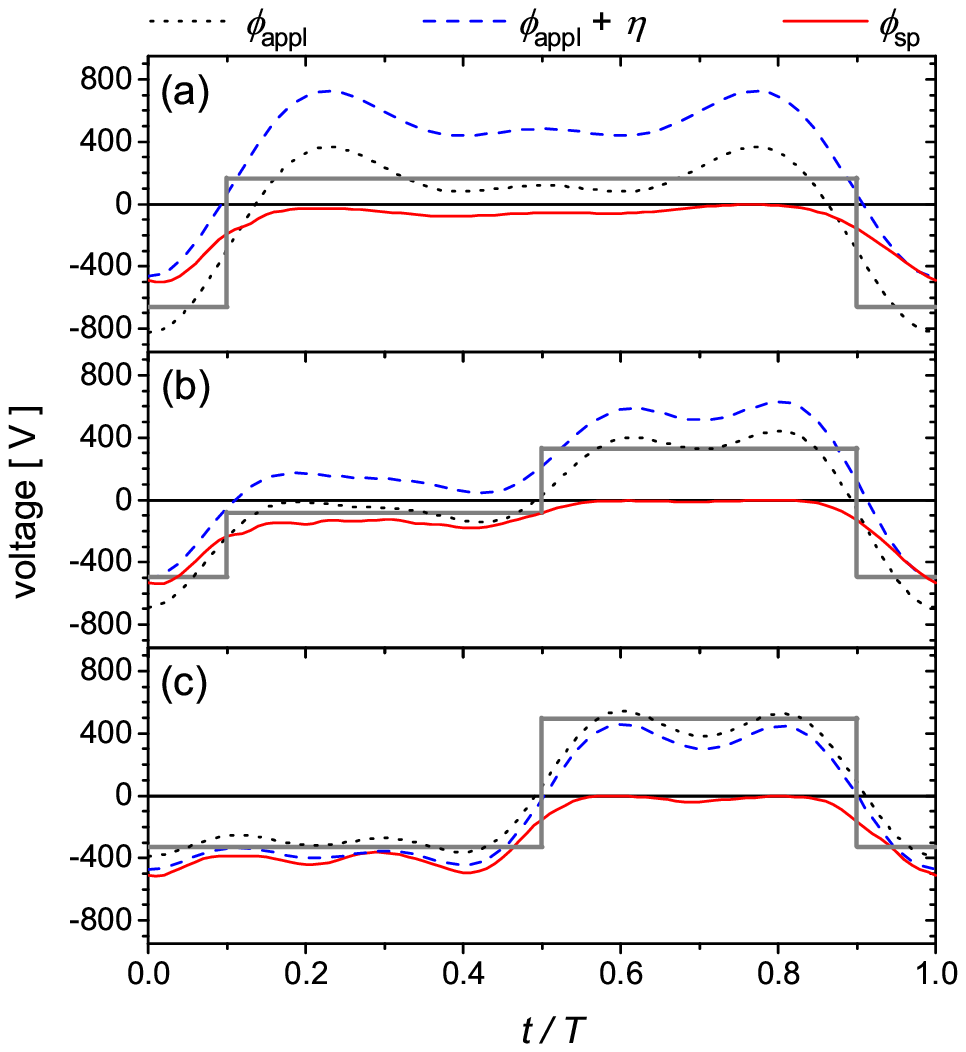}
\caption{Driving voltage waveform ($\phi_{\rm appl}$), discharge voltage including DC self-bias ($\phi_{\rm appl} + \eta$), and voltage drop over the powered electrode sheath ($\phi_{\rm sp}$), for voltage plateau values of (a) 0 \%, (b) 50 \%, and (c) 100 \%, i.e. voltage parameters specified in Table \ref{table1}. The grey lines indicate the desired applied voltage waveform.}
\label{fig:6}
\end{center}
\end{figure}

\begin{table}
\caption{Amplitudes and phases of the applied harmonics and DC self-bias, $\eta$, for different plateau values (PVs).}
\label{table1}
\begin{indented}
\item[]
\begin{tabular}{@{}l*{20}{c}}
\br
 PV \hspace{2 mm} & $\phi_1$ & $\theta_1$ & $\phi_2$ & $\theta_2$ & $\phi_3$ & $\theta_3$ & $\phi_4$ & $\theta_4$ & $\phi_5$ & $\theta_5$ & $\eta$ \\
 $\textnormal{[\%]}$                 & [V]           & [$^\circ$]   & [V]           & [$^\circ$]   & [V]           & [$^\circ$]   & [V]           & [$^\circ$]   & [V]           & [$^\circ$]   & [V]    \\
\mr
 0    & 265 & 180  & 310 & 180  & 154 & 180  & 80.9 & 180  & 14.8 & 180  & 414  \\
 10  & 268 & 170  & 307 & 182  & 154 & 178  & 80.4 & 188  & 15.5 & 180  & 396  \\
 20  & 276 & 160  & 300 & 184  & 152 & 175  & 80.4 & 196  & 16.1 & 180  & 363  \\
 30  & 290 & 150  & 289 & 187  & 149 & 172  & 81.2 & 205  & 16.4 & 180  & 317  \\
 40  & 309 & 141  & 273 & 190  & 144 & 169  & 82.8 & 213  & 16.6 & 180  & 257  \\
 50  & 331 & 133  & 255 & 193  & 137 & 166  & 85.5 & 222  & 16.6 & 180  & 186  \\
 60  & 355 & 126  & 234 & 197  & 131 & 162  & 89.0 & 230  & 16.4 & 180  & 115  \\
 70  & 381 & 121  & 212 & 201  & 124 & 158  & 93.1 & 236  & 16.1 & 180  & 46.0  \\
 80  & 406 & 116  & 189 & 205  & 117 & 154  & 97.6  & 242  & 15.7 & 180  & -22.8  \\
 90  & 431 & 112  & 167 & 210  & 110 & 149  & 102   & 248  & 15.3 & 180  & -63.2  \\
 100& 454 & 108  & 145 & 216  & 104 & 144  &107   & 252  & 14.8 & 180  & -82.0  \\
\br
\end{tabular}
\end{indented}
\end{table}

Next, we demonstrate the effect of modifying the shape of the applied voltage waveform. The panels of figure \ref{fig:6} display three cases. The ``ideal'' driving voltage waveforms, $\phi_{id}(t)$, are shown by the grey lines. The amplitudes and phases of the individual frequency components are obtained by a Fourier transform of the desired voltage waveform. The black dotted curves show the approximation of the desired waveforms by 5 Fourier components, which is used in the simulations. The amplitudes and phases of these Fourier components are provided in table \ref{table1}.

The ideal waveform consists of three plateaus with values $V_1$ between $t=-0.1$ $T$ ($0.9$ $T$) and $t=0.1$ $T$, $V_2$ between $t=0.1$ $T$ and $t=0.5$ $T$, and $V_3$ between $t=0.5$ $T$ and $t=0.9$ $T$ ($V_1 \leq V_2 \leq V_3$). Furthermore, the voltage waveforms have no DC component, i.e. $\int_0^T \phi_{id}(t) dt =0$. Here, $V_2$ is varied between $V_3$ and $V_1$, which are kept at a constant difference. The plateau values (PV) of  0 \% and 100 \% correspond to $V_2=V_3$ and $V_2=V_1$, respectively.
The voltage amplitudes and phases of the individual frequency components of the applied voltage (equation (\ref{AppVol}) with $A_0=1.0$) are provided in table \ref{table1}. This variation causes a change not only in the overall shape of the applied voltage waveform, but also in the asymmetry of the discharge, so that the DC self-bias, $\eta$ (also included in table \ref{table1}), changes as a function of the PV, as well. 

\begin{figure}[t]
\begin{center}
\includegraphics[width=0.75\textwidth]{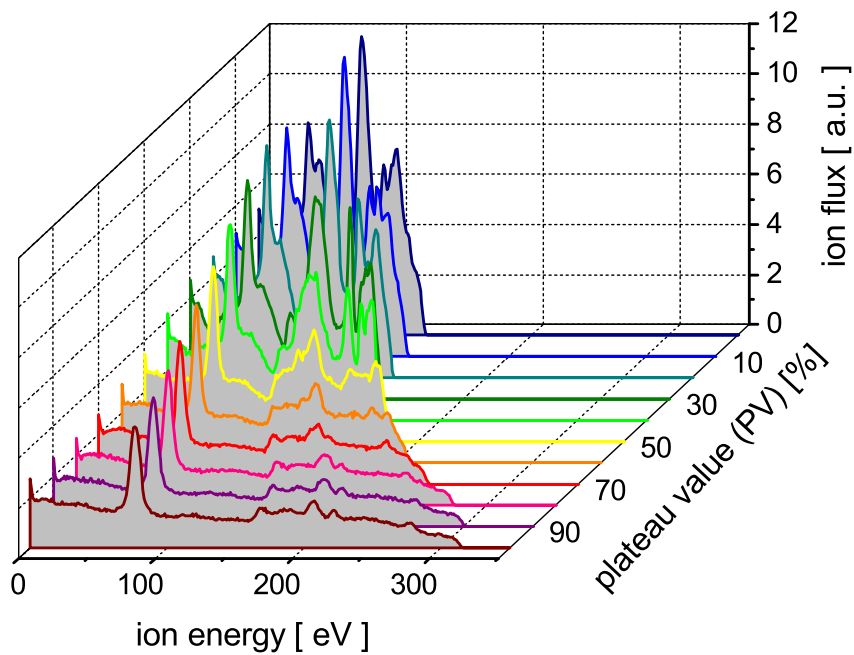}
\caption{Flux-energy distribution functions of He$^+$ ions at the powered electrode, for different plateau values (PVs) of the driving voltage waveform.}
\label{fig:7}
\end{center}
\end{figure}

The resulting IDFs for different PV are depicted in figure \ref{fig:7}. For a PV of 0 \%, the IDF is very narrow, with a maximum ion energy of about 124 eV. Increasing the PV leads to an increase of the voltage drop across the powered electrode sheath and, hence, to an increase of the IDF width to up to 318 eV. Meanwhile, the peak in the IDF due to cx-ions, i.e. caused by the physical mechanisms discussed above, is present for all PVs, but it becomes more narrow for larger PV. It is one of multiple peaks at PVs below $\approx$ 30 \%, whereas it is the dominant feature for PVs larger than $\approx$ 30 \%. Above this plateau value, the total flux within the width of the peak stays almost constant (it varies by $\pm17$ \%, not shown), whereas the energy of the peak can be tuned by almost a factor of two (between 43 eV and 78 eV) by ajusting the PV from 30 \% to 100 \%. 

At small PVs (e.g. 0\%), the long time interval of a small sheath voltage and, hence, of a small electric field leads to many cx-ions accumulating in the sheath region at the powered electrode. However, only a small fraction of these ions arrive at the electrode after a sufficiently short transit time. Thus, the collection of cx-ions within a transit time of one RF period or less is very inefficient. With an intermediate PV (e.g. 50 \%), this collection is improved, but the longer time interval of a significant sheath electric field also means a shorter accumulation phase. A high PV (e.g. 100 \%) allows for a larger energy gain of cx-ions, so that the respective peak in the IDF is shifted towards higher energies. Meanwhile, the relative contribution of the flux of the ions within the peak to the total ion flux slightly decreases, from 17.2 \% at PV= 50 \% to 16.2 \% at PV= 100 \%.

In both variations shown in figures \ref{fig:2}  and \ref{fig:7}, the main peak is superimposed on a broad spectrum of the IDF. A further increase of the relative flux of the dominant peak would be worthwhile and might be achievable by changing more parameters, such as the gas pressure or the fundamental driving frequency. Such a wide parameter variation is beyond the scope of this work, though. Furthermore, there are further peaks at higher energies, which are caused by cx ions that are created further away from the electrode and have a transit time of multiple RF periods. Therefore, the development of multiple peaks is inherently linked to the periodic behavior of the RF sheath \cite{IEDF3,IEDF4,IEDF5}. Nevertheless, these peaks are much less pronounced, because the spatial regions, in which cx ions may accumulate, become narrow far away from the electrode (see figure \ref{fig:5}) and because the probability of collisions, in which the energy of these ions is redistributed, becomes larger. Certainly, the application of the approach discussed here is limited to gases with a symmetrical charge exchange between the primary ion species and the feed gas.

\section{Conclusions}

In conclusion, in this work we presented a novel method to generate a distinct peak at low/intermediate energies in the flux-energy distribution of the ions impinging on the electrode surfaces in capacitive multi-frequency plasmas. This was made possible by driving the plasma with a tailored voltage waveform that consists of five consecutive harmonics. Subject to the Electrical Asymmetry Effect, a DC self-bias develops as a function of the phases between the applied harmonics. We have shown that by adjusting the applied voltage waveform to approximate certain idealized waveforms, the overall shape of the IDF, as well as distinct features within the distribution function can be controlled.
We found that utilizing the effect of charge exchange collisions of the ions in the sheaths, a dominant peak in the IDF can be created. 
This approach works via a detailed knowledge of the ion dynamics within the sheath and is, in particular, based on a control of the dynamics of cx-ions in the spatio-temporal sheath electric field. The general idea of dividing the RF period into intervals of cx-ion accumulation (small sheath electric field) and collection (high and intermediate sheath electric field) was tested and the resulting peak in the IDF was analyzed. It was found that the position (i.e. the energy) and the flux of the peak can be varied at the same time by adjusting the amplitude of the applied voltage waveform, and can be controlled almost independently of one another by changing the PV of the intermediate voltage plateau. These advanced control opportunities allow the generation and control of a single peak at low/medium energies within the IDF, so that certain energy-selective reaction process at the substrate surface induced by ion bombardment within a distinct energy window can be specifically enhanced, while the risk of inducing other potentially parasitic reactions initiated by ions at different energies is reduced. 
Certainly, this study is a proof of principle and the method needs to be tested, e.g. for other feed gases, in a future study.

\ack This work was supported by the Hungarian Fund for Scientific Research (OTKA), via grants K105467 and NN-103150.


\end{document}